\documentclass[preprint,superscriptaddress,prl,longbibliography]{revtex4-1}
\usepackage{graphicx}
\usepackage{amsmath,amssymb}
\usepackage{mathrsfs}
\usepackage[normalem]{ulem}
\usepackage{color}
\usepackage{soul}
\usepackage{hyperref}
\usepackage{textcomp}

\usepackage[resetlabels, labeled]{multibib}

\usepackage{natbib}

\usepackage[T1]{fontenc}
\usepackage[utf8x]{inputenc}
\usepackage{bm}
\usepackage[left]{lineno}
\linenumbers

\newcommand{\be}{\begin{equation}}
\newcommand{\ee}{\end{equation}}
\newcommand{\bea}{\begin{eqnarray}}
\newcommand{\eea}{\end{eqnarray}}

\newcommand{\ignore}[1]{}

\newcommand{\UFMGPPGEE}{Programa de Pós-Graduação em Engenharia Elétrica, Universidade Federal de Minas Gerais, Belo Horizonte, MG 31270-901, Brazil}
\newcommand{\UFMGDELT}{Departamento de Engenharia Eletrônica, Universidade Federal de Minas Gerais, Belo Horizonte, MG 31270-901, Brazil}
\newcommand{\UFMGPPGF}{Programa de Pós-Graduação em Física, Universidade Federal de Minas Gerais, Belo Horizonte, MG 31270-901, Brasil.}
\newcommand{\UFMGLCPN}{Laboratório de Caracterização e Processamento de Nanomateriais (LCPNano), Universidade Federal de Minas Gerais, Belo Horizonte, MG 31270-901, Brazil.}
\newcommand{\UFMGDF}{Departamento de Física, Instituto de Ciências Exatas, Universidade Federal de Minas Gerais, Belo Horizonte, MG 31270-901, Brazil.}
\newcommand{\UIL}{Frederick Seitz Materials Research Laboratory, University of Illinois, 104 S. Goodwin Avenue, Urbana, IL 61801, USA.}
\newcommand{\CDTN}{Laboratório de Química de Nanoestruturas de Carbono, Centro de Desenvolvimento da Tecnologia Nuclear, Belo Horizonte, MG 31270-901, Brazil.}
\newcommand{\UFU}{Instituto de Ciências Exatas e Naturais do Pontal, Universidade Federal de Uberlândia, Ituiutaba, MG 38304-402, Brazil.}
\newcommand{\UFMGICB}{Departamento de Bioquímica e Imunologia, Universidade Federal de Minas Gerais, Belo Horizonte, MG 31270-901, Brazil.}
\newcommand{\Inatel}{Instituto Nacional de Telecomunicações, Santa Rita do Sapucaí, MG 37540-000, Brazil.}

\renewcommand{\phi}{\varphi}
\renewcommand{\epsilon}{\varepsilon}

\usepackage{siunitx}

\begin{document}
\nolinenumbers

\title{Tunable Surface Plasmon-Polaritons Interaction in All-Metal Pyramidal Metasurfaces: Unveiling Principles and Significance for Biosensing Applications}

\author{Talles E. M. Marques}
\affiliation{\UFMGDELT}

\author{Yuri H. Isayama}
\affiliation{\UFMGLCPN}
\affiliation{\UFMGDF}

\author{Felipe M. F. Teixeira}
\affiliation{\UFMGDELT}
\affiliation{\UFMGPPGEE}

\author{Fabiano C. Santana}
\affiliation{\UFMGPPGF}

\author{Rafael S. Gonçalves}
\affiliation{\UFMGLCPN}
\affiliation{\UIL}

\author{Aline Rocha}
\affiliation{\UFMGICB}

\author{Bruna P. Dias}
\affiliation{\CDTN}

\author{Lidia M. Andrade}
\affiliation{\UFMGDF}

\author{Estefânia M. N. Martins}
\affiliation{\CDTN}

\author{Ronaldo A. P. Nagem}
\affiliation{\UFMGICB}

\author{Clascidia A. Furtado}
\affiliation{\CDTN}

\author{Miguel A. G. Balanta}
\affiliation{\UFU}

\author{Jorge Ricardo Mej\'ia-Salazar}
\affiliation{\Inatel}

\author{Paulo S. S. Guimarães}
\affiliation{\UFMGPPGF}
\affiliation{\UFMGDF}

\author{Wagner N. Rodrigues}
\affiliation{\UFMGPPGF}
\affiliation{\UFMGDF}

\author{Jhonattan C. Ramirez *,}
\affiliation{\UFMGDELT}
\affiliation{\UFMGPPGEE}

\maketitle

\textbf{Abstract} 

The strong coupling of plasmonic resonance modes in conductive pyramidal nanoparticles leads to an increase in the density of free charges on the surface. By ensuring plasmonic coupling in the pyramidal nanoparticle lattice, the achieved field intensity is potentiated. At the same time, a strong coupling between resonant modes is guaranteed, which results in the formation of new hybrid modes. In this manuscript, we demonstrated a tunable double anticrossing interaction that results from the interaction between two Localized Surface Plasmon Resonance (LSPR) modes and a Surface Plasmon Polariton (SPP) wave. The tuning is done as a function of the variation of the angle of incidence of the input electric field. From the double anticrossing, an increase in field intensity in a blue-shifted LSPR mode located in the red wavelength region is observed. This demonstrates that at certain angles of incidence, the intensity field obtained is strongly favored, which would be beneficial for applications such as Surface Enhancement Raman Spectroscopy (SERS). Nanoparticle-based lattices have been widely used for biosensor applications. However, one of the major limitations of this type of device is the low tolerance to high concentrations of biomolecules, which significantly affects their performance. According to the studies carried out for this manuscript, it was demonstrated that the implemented geometry allows for the observation of an LSPR mode, which is responsible for the control and synchronization of other perceived resonances. This mode remains almost invariant when subjected to structural variations or changes in the angle of incidence of the electric field. These characteristics eliminate the limitation mentioned above, allowing for sensitivities $10^3$ times higher than those achieved in conventional systems based on LSPR used to detect \textit{P. brasiliensis} antigen. This type of fungus is responsible for causing Paracoccidioidomycosis (PCM), which is a systemic granulomatous mycosis. The implemented geometry also guarantees high tolerance to high concentrations, as the latter affects the resonance on the surface, but not the localized modes, as demonstrated.

\section{Introduction}

Plasmonic nanostructures are pivotal in recent transformative breakthroughs spanning biosensing~\cite{Wang2020,Kong2020,Luo2021}, waveguiding~\cite{Ono2020}, imaging~\cite{Okamoto2022}, energy harvesting~\cite{Dhiman2020,Lee2021}, and beyond~\cite{Rizal2022}. Rooted in surface plasmon polaritons, these nanostructures exploit the resonant coupling of optical fields with surface charge density oscillations on metal surfaces, enabling the confinement, enhancement, and localization of light at subwavelength scales. Such resonances foster innovation across diverse domains, facilitating the development of high-performance optical devices that surpass the diffraction limit~\cite{Gramotnev2010}, thus enabling seamless on-chip integrability. In biosensing, for instance, the synergy between plasmonic nanostructures and microfluidic platforms, combined with their heightened sensitivity to changes of dielectric properties near the surface (induced through adsorption processes), holds promise for the burgeoning field of point-of-care (PoC) diagnostic devices~\cite{MejiaSalazar2018}. However, despite significant progress, persistent challenges hinder the transition of plasmonic biosensors from research laboratories to handheld devices. In SERS, for example, a notable challenge arises from the limited signal amplification primarily due to the inhibited interaction between the electromagnetic field in the dielectric substrate and analyte molecules. Moreover, traditional plasmonic biosensing methods encounter limitations due to relatively modest electromagnetic field enhancements in flat planar surfaces or grating structures~\cite{Hilal2022,Zhang2023}. Addressing these challenges requires innovative methodologies to engineer highly efficient plasmonic devices capable of maximizing electromagnetic field concentration at the analyte region~\cite{MejiaSalazar2018,ehsan2023,zhou2023}. Furthermore, these nanostructures must seamlessly integrate with microfluidic technology to ensure easy implementation in future PoC applications. On the other hand, the flourishing field of optical metasurfaces is attracting growing interest due to its remarkable ability to precisely manipulate and adapt optical properties~\cite{Neshev2018,Lin2021,Du2022,Pertsch2023}. These metasurfaces harness the near-field overlap between neighboring nanoparticles, mimicking the electronic bands found in well-localized atomic orbitals. Within the domain of plasmonic nanoresonators, this phenomenon --often termed as plasmon hybridization-- enables tailored adjustments to optical behavior, thus facilitating the creation of distinctive optical features through meticulous design of the elementary plasmonic constituents~\cite{Vaekevaeinen2014,yang2018,Kravets2018,Oleynik2024}. Indeed, groundbreaking achievements have been demonstrated in spatial resolution levels, particularly in light's polarization control, holograms, and wavefront manipulation. Additionally, unprecedented efficiencies have been achieved in bending and focusing light~\cite{Kuznetsov2024}. Nevertheless, these metasurfaces predominantly rely on metal nanoparticles positioned on dielectric substrates~\cite{Kravets2018} and/or the utilization of hybrid metal-dielectric nanostructures~\cite{kim2023,chu2023}, which consequently diminishes the electromagnetic field directly interacting with the analyte region.

Similarly, the plasmonic interactions, where lattice-based nanostructures exhibit anticrossing behavior due to the interaction generated between propagated waves and localized fields tuned by the variation in the size and separation of nanoparticles, has been of great value \cite{ghoshal2008,ghoshal2009,yang2018}. These effects offer advantages in modern communication systems, biological sensing systems, alternative energy generation, and more. However, the generation of these effects in phonon-polaritons \cite{zubin2014,gubbin2019,chafatinos2020,kachiraju2022}, exciton-polaritons \cite{gibbs2011}, and plasmon-polaritons \cite{ghoshal2008,ghoshal2009,garciadeabajo2021} systems is also observed in metal/dielectric systems.

In light of these challenges, our study aims to address the limitations of existing nanostructures for plasmonic biosensing. Inspired by remarkable achievements in spatially coherent tip-enhanced Raman spectroscopy measurements utilizing a single pyramidal structure~\cite{nadas2023,olJhonattan2024}, we hypothesized that metasurfaces composed of two-dimensional arrays of gold nanopyramids could serve as an ideal substrate for plasmonic biosensing. 

To surmount the limitations linked with dielectric substrates, we investigate the strong coupling between modes generated through the plasmon-polariton interaction in an all-metal plasmonic metasurface. The spectral response is tuned by varying the angle of incidence of the light beam, facilitated by the triangular shape of the pyramidal nanoparticles. Double-anticrossing behavior is observed in the acquired absorption spectrum, resulting from the strong coupling between the LSPR mode and the generated surface plasmon polariton (SPP) wave. Furthermore, it has been observed a blue-shifted strong LSPR mode occurring between 600 nm and 700 nm wavelength, which can be tuned by varying the angle of incidence of the electric field, leading to an increase in the collective oscillations in the pyramid-based nanoparticle lattice. This strong interaction suggests that this pyramidal geometry, combined with the angle of incidence of the electric field, significantly favors effects such as Raman spectroscopy at this wavelength. 

We embraced a cost-effective and widely accessible lithography technique to produce a silicon mask featuring a two-dimensional array of pyramids. The subsequent fabrication of an all-metal plasmonic metasurface involved the deposition of a 175 nm-thick gold layer onto the silicon mask. This specific thickness was meticulously selected to prevent any electromagnetic interaction with the substrate, thus mitigating potential drawbacks associated with dielectric substrates. Moreover, our innovative prism-fabrication methodology produces high-quality pyramids, as it is demonstrated with several morphological characterizations of the samples. The process previously described allows the mass manufacturing of these plasmonic metasurfaces. In contrast to the complexity of utilizing single pyramidal nanoparticles for successful measurements, the metasurface developed in this study offers a straightforward integration into portable and compact  devices, making it suitable for PoC deployment~\cite{Ngo2022}. The suitability of our approach for biosensing applications underwent validation through two distinct methods. Initially, SERS measurements were conducted using urea as the target analyte. Subsequently, the two-dimensional configuration of gold nanopyramids demonstrated remarkable potential in exciting localized SPRs. This property facilitated the monitoring of Pb27r protein presence (derived from \textit{Paracoccidioides brasiliensis}) by observing shifts in the LSPR resonant wavelength. Notably, this study achieved detection levels of up to $\mu$g/mL, marking the inaugural application of plasmonic biosensing for Pb27r protein label-free detection. The simplicity and high sensitivity of our system are noteworthy. The gold-based surface is easily functionalized using conventional techniques, and its structure aligns seamlessly with flow-over sensing devices, readily integrable with microfluidic channels. Consequently, our proposal sets the stage for future PoC devices capable of multiplexed analyses within a single plasmonic platform.

\section{Results and discussions}
\subsection{LSPR on pyramidal-based nanoparticle lattice}

Recently, the implementation of pyramidal structures has exhibited outstanding performance in amplifying Raman responses for analyzing diverse materials. This approach has proven particularly valuable for investigating the electro-optical properties of complex materials, such as twisted bilayer graphene \cite{natureAdo,nadas2023}. The efficiency of the pyramidal-based tip in Raman spectroscopy amplification has been demonstrated, attributed to its ability to strongly confine the electric field, thereby generating a well-established electro-optical interaction \cite{hudson2020, olJhonattan2024}.

The same operating principle used in Tip-Enhancement Raman Spectroscopy (TERS) systems is applied to devices that rely on conductive pyramidal structures. However, a periodic system needs to be defined for this type of component, which is not conventionally considered in various experimental demonstrations found in the literature \cite{roy2017,li2020,thuy2021,chang2022,sousa2023}. However, it is also important to highlight that some works have taken advantage of the geometric characteristics and periodicity of the pyramids to improve Raman signal enhancement for biodetection processes \cite{liu2022, das2023}. 

\begin{figure}[!hbtp]
\centering
\centerline{\includegraphics[width=\textwidth]{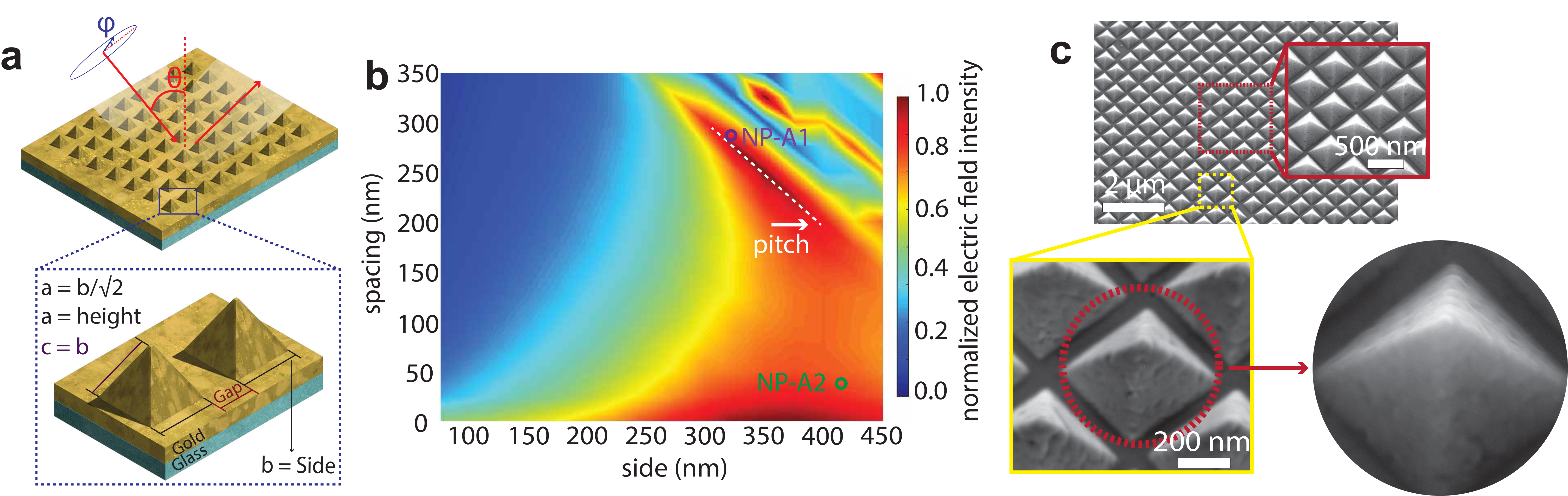}}
\caption{\textbf{Pyramidal nanoparticle lattice device.} \textbf{a} Schematic diagram representing the numerically modeled pyramidal nanoparticle lattice: \textit{a} represents the height, \textit{b} is the base, and \textit{c} is the edge of the pyramids. $\theta$ represents the angle of incidence of the interacting light beam, and $\phi$ is the azimuthal angle. \textbf{b} The normalized electric field intensity is calculated based on the variations in the pyramids' dimensions to determine the most appropriate combination of parameters that can help achieve the highest field intensity. \textbf{c} High-quality pyramidal nanoparticle lattice has been successfully manufactured.}
\label{fig1}
\end{figure}

By adjusting parameters such as the size of the base and the separation of elements in a lattice of pyramidal nanoparticles, we can ensure plasmonic coupling between them, creating a symmetrical x-spaced structure, like the one shown in Fig.~\ref{fig1}a. This guarantees that the set of elements behaves like a single device, resulting in strong electric field confinement, facilitating electron movement on the surface, as demonstrated in Fig.~\ref{fig1}b. However, this strong confinement is restricted to a specific set of combinations involving the base size of the pyramid and the spacing between elements in the array, termed the "Iso-pitch" region, as presented in the numerical simulations illustrated in Fig.~\ref{fig1}b. It is crucial to note that while ensuring optimal performance in the proposed plasmonic system, we simultaneously provide a higher tolerance for errors introduced during the manufacturing of nanodevices. The characteristics of the modeled and subsequently manufactured pyramids are presented in the inset of the Fig.~\ref{fig1}a.

Fig.~\ref{fig1}b highlights the NP-A1 device (330 nm side – 290 nm spacing) and the NP-A2 device (420 nm side – 40 nm spacing), which refer to pyramid arrays manufactured as shown in Fig.~\ref{fig1}c, following the established reference in Fig.~\ref{fig1}b, and the methodology outlined in both the methods section and Fig. S1 of the supplementary material. 

\subsection{Strong coupling between LSPR modes and SPP waves}

Pyramidal structures favor the achievement of more intense electric fields than other types of in-plane nanoparticles, such as nanodiscs and nanoblocks \cite{das2023}. This is due to its geometric symmetry and sharp vertices. However, as will be presented later, we demonstrated that the absorption achieved can significantly increase depending on the angle of incidence of the light beam. This is because the interaction between an incident electric field and a surface with an elevation angle $\theta$ induces a more significant movement of charges, thereby intensifying the electric field in the pyramidal nanoparticle lattice. Preliminary simulations presented in Fig. S2 (supplementary material) can provide early evidence for this phenomenon.

\begin{figure}[!hbtp]
\centering
\centerline{\includegraphics[width=\textwidth]{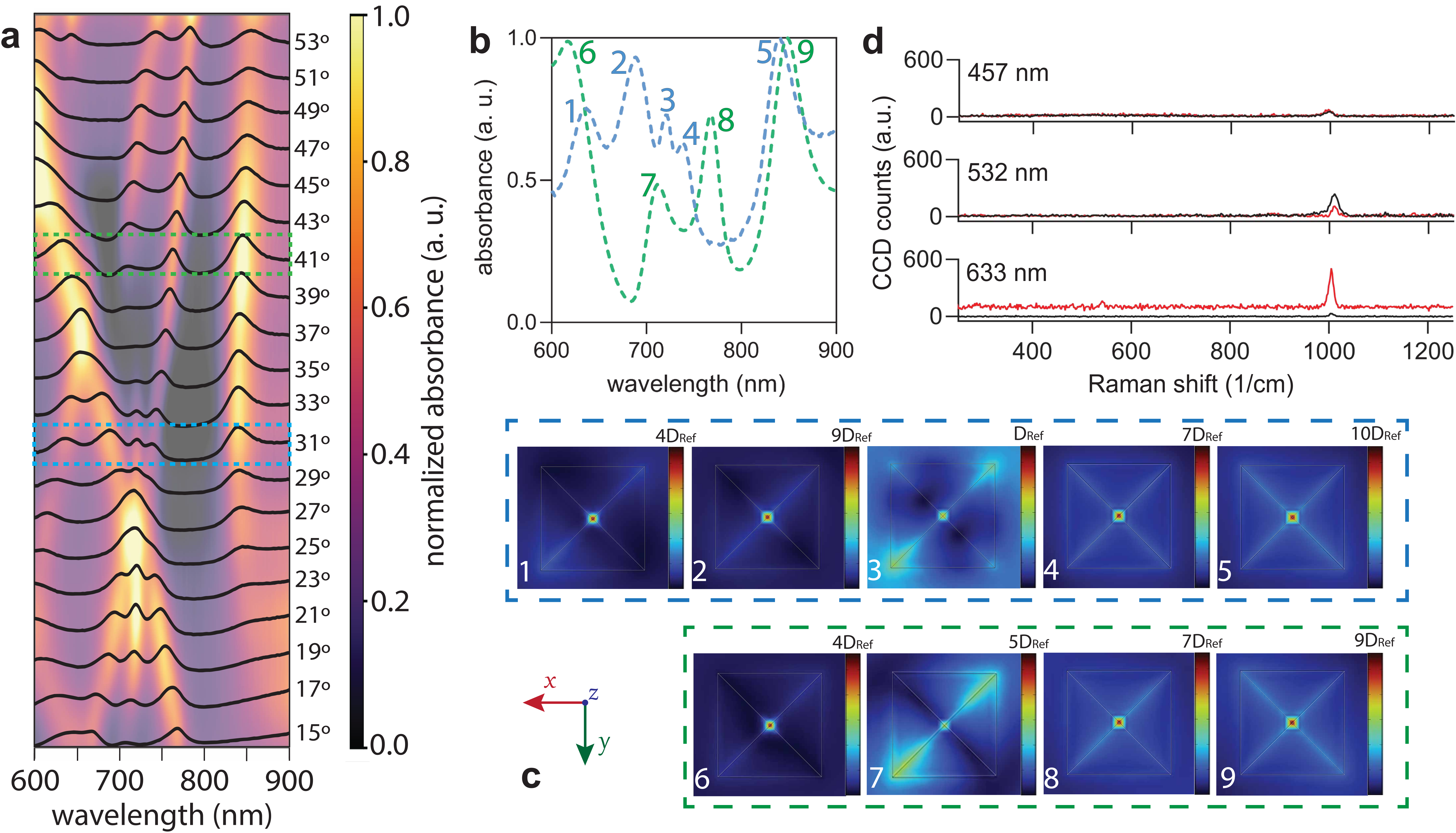}}
\caption{\textbf{Tunable surface plasmon-polaritons coupling.} \textbf{a} Measured spectral response for the pyramidal nanoparticle lattice shows a tunable surface plasmon-polariton coupling interaction that is highly dependent on the localized surface plasmonic resonance at a wavelength of 850 nm. Additionally, a double anticrossing phenomenon is observed near 725 nm wavelength, which is a result of the interaction between two LSPR modes and one SPP wave. Moreover, strong confinement is noticed in the blue-shifted LSPR mode, which progressively increases in energy with an increase in the angle of incidence near 633 nm wavelength. \textbf{b} Measured spectral response of the LSPR modes and SPP wave for incidence angles of \ang{31} (blue dashed line) and \ang{41} (green dashed line). \textbf{c} Simulated electric displacement field distribution for each peak numbered in \textbf{b}. \textbf{d} Raman spectra for 40 mMol/L of urea on our pyramidal nanoparticle lattice substrate. The Raman response was measured for three different wavelengths 457 nm, 532 nm, and 633 nm. The black line represents the measurement made on the sample deposited on the smooth metal surface, the red line represents the measurement made on the pyramidal nanoparticle lattice.}
\label{fig2}
\end{figure}

Fig.~\ref{fig2}a shows the spectral response for different measured angles. This result is possible only because the dimensions chosen for our device satisfy the conditions established in Fig.~\ref{fig1}b (Device NP-A1). This device is close to the "Iso-pitch" region, so distinct and well-defined plasmonic resonances can be observed. On the other hand, no resonance will be observed in regions where the electric field has low intensity. This behavior was further verified with the manufacturing and testing of the NP-A2 devices, as shown in Fig. S3a of the supplementary material, which did not exhibit any significant localized plasmon resonance in any region of the spectrum evaluated, regardless of the angle of incidence of the electric field. The high quality of the array of the manufactured devices, which corresponds to the NP-A1 device from Fig.~\ref{fig1}b, can be corroborated in Fig. S3b, as well as in the supplementary material, where we present the measurements of the size from the manufactured devices.

The absorption spectra were analyzed for different angles of incidence, ranging from \ang{15} to \ang{53}. The results of this analysis can be appreciated in Fig.~\ref{fig2}a. For angles of incidence below \ang{15}, a single peak was observed within the 750 nm and 800 nm wavelength range. However, with the emergence of an additional peak at wavelengths near 850 nm, three distinct energy states are now visible between 650 nm and 800 nm wavelengths. With variations in the angle of incidence, it can be observed that the central peak near 725 nm, remains almost unshifted; however, the peaks on the left and right exhibit a significant variation in wavelength depending on the incident electric field's variation.

The central peak corresponds to the SPP wave, while the lateral peaks are associated with LSPR modes. Their interaction in the described region configures a double-anticrossing between the LSPR modes and the SPP wave. In Fig.~\ref{fig2}b, the interacting peaks are prominent at an angle of incidence of \ang{31}, and its field distribution can be seen in Fig.~\ref{fig2}c. The SPP wave can be observed in Fig.~\ref{fig2}c-3 through the electric displacement field distribution corresponding to peak three in Fig.~\ref{fig2}b. The remaining peaks correspond to well-defined LSPR modes.

It is worth noting that the SPP wave vanishes as the angle of incidence surpasses \ang{37}. Nevertheless, it is coupled with an LSPR mode, which arises when the angle of incidence reaches \ang{39}. This mode is visible in peak seven of Fig.~\ref{fig2}b, where the electric displacement field distribution depicted in Fig.~\ref{fig2}c-7 exhibits greater confinement with surface plasmon remnants as expected.

From the double-anticrossing explained earlier, we can observe a blue-shifted LSPR mode that progressively increases in energy with an increase in the angle of incidence near 633 nm wavelength. This mode collides with another LSPR mode, absorbing it entirely. The observed increase in energy of this LSPR mode is due to the variation of the angle of incidence of light, which allows for tuning of the density of free charges on the surface of the pyramidal nanoparticle lattice. The response of this nanopyramid system favors other types of applications, such as Surface Enhancement Raman Spectroscopy (SERS), which has demonstrated a significant increase in enhancement, up to 7x compared to a flat gold surface, as can be seen in Fig.~\ref{fig2}d, setting the laser at 633 nm. No significant enhancements were observed in the other cases when the laser was set at 457 nm and 532 nm wavelengths. At 457 nm, a similar response was observed in the Raman signal obtained inside and outside the pyramid region. However, at 532 nm, a slightly higher enhancement was observed in the region without the pyramids. This observation is consistent with the region where gold exhibits greater absorption.

We conducted numerical simulations to analyze the maximum electric field intensity achieved as a function of the incident electric field intensity $(|E|/|E_{in}|)$, for a flat gold surface and pyramidal nanoparticle lattice at \ang{0} and \ang{40} angles of the incident. Based on the obtained results, we observed an enhancement of 0.61x for the flat gold surface, which is expected since the incident electric field will scatter on the surface at \ang{0}. However, for the pyramidal nanoparticle lattice at \ang{0} and \ang{40}, we observed an enhancement of 2.62x and 70.72x, respectively. The previous suggests that a SERS system excited at a certain angle of incidence would present an extremely improved performance.

It is worth noting that the LSPR mode at 850 nm achieves its highest intensity at an angle of \ang{37}. This mode exhibits a nearly constant absorption wavelength, regardless of the angle of incidence of the electric field. Additionally, it enables the control of the behavior of the other resonant states towards the incoming radiation, as demonstrated.

\subsection{Impact on biosensing applications.}

Based on the angle of incidence of the electric field during its interaction with our periodic system of pyramidal nanoparticles, we were able to observe that the LSPR mode exhibiting the greatest stability to structural changes and modifications was the peak observed at wavelengths near 850 nm. The stability demonstrated by the LSPR mode mentioned above makes this device a viable option for sensing applications because it ensures that the changes observed on the device's surface are a result of external elements' variations in a specific way.

In order to verify the device's performance as a biosensor, the protein Pb27r from \textit{P. brasiliensis} was produced and subsequently detected in varying concentrations. Paracoccidioides (\textit{P. brasiliensis}) is a type of dimorphic fungus responsible for causing Paracoccidioidomycosis (PCM), which is a systemic granulomatous mycosis. This disease is endemic to Latin America and is considered one of the ten leading causes of death due to infectious and parasitic diseases, chronic and recurring, in Brazil. The methods section details the process of protein production and sample preparation for the various concentrations used in the experiments.

\begin{figure}[!hbtp]
\centering
\centerline{\includegraphics[width=0.4\textwidth]{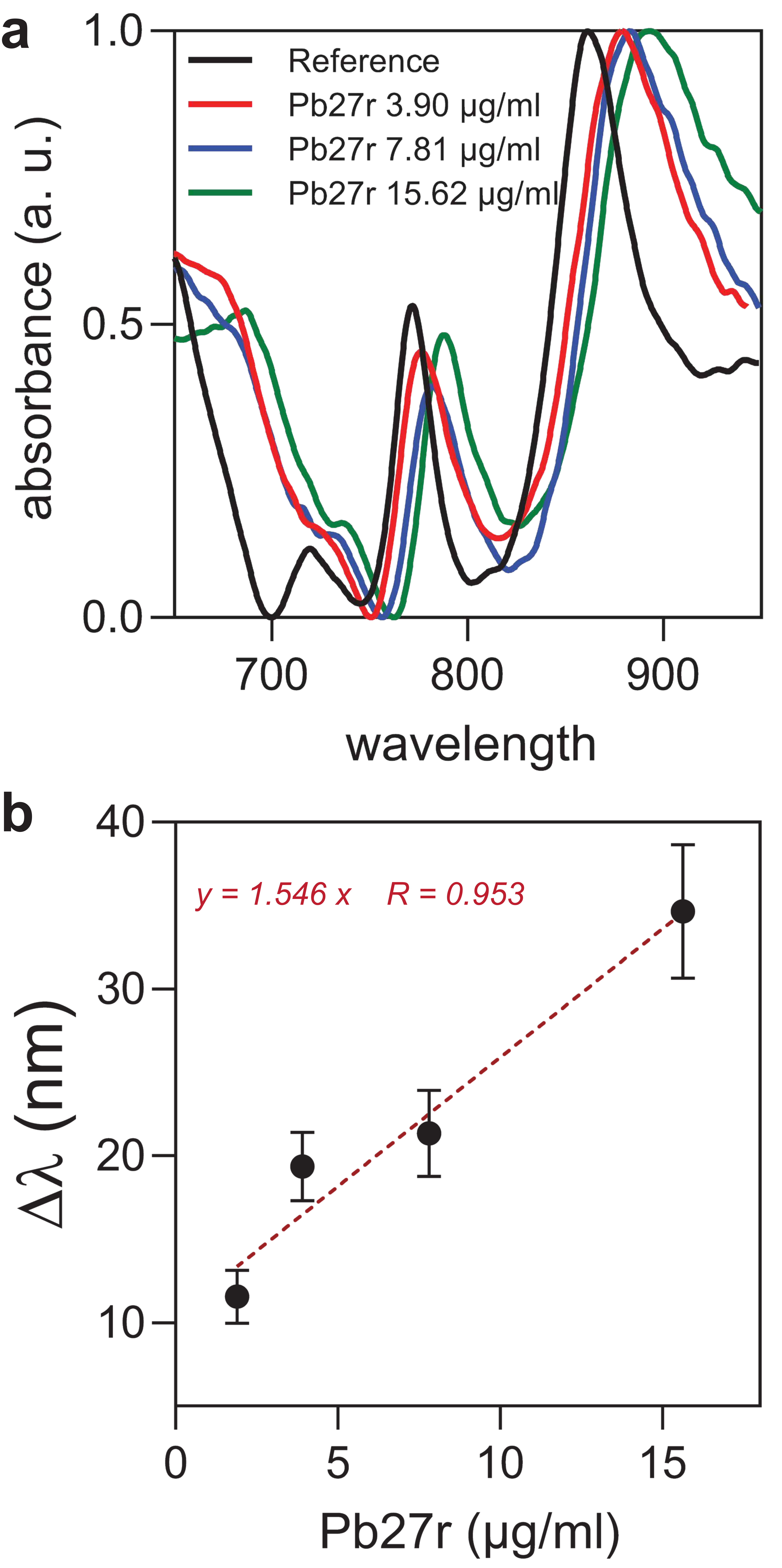}}
\caption{\textbf{Detection of the Pb27r protein.} \textbf{a} Response of the plasmonic resonance of the pyramidal nanoparticle lattice, demonstrating the shift in wavelength of the curve measured for various concentrations of the Pb27r protein. \textbf{b} The linear calibration curve for various concentrations of Pb27r.}
\label{fig3}
\end{figure}

The results shown in Fig.~\ref{fig3}a clearly demonstrates a shift in the plasmonic resonance curve with an increase in the concentration of Pb27r. This phenomenon is expected since the concentration of protein on the surface of our device increases as well.

We conducted experiments with concentrations ranging from 1.9 $\mu$g/ml to 15.62 $\mu$g/ml. The results are shown in Fig.~\ref{fig3}b, which allowed us to observe that our device has a sensitivity of 1.546 nm/$\mu$g/ml. It is worth noting that conventional LSPR systems typically achieve sensitivities of the order of mg/ml for this particular protein. This means that our device is $10^3$ times more sensitive, which can be attributed to the high density of free charge we observed.

SERS systems allow accurate detection of molecules with known Raman response. However, this becomes a hurdle in detecting challenging diseases that require rapid response times and high precision, such as SARS-CoV-2. Our proposal offers a solution to this problem by allowing the detection of specific proteins through label-free systems that rely on the antigen-antibody interaction or DNA/RNA association. This approach guarantees an excellent spectral response with high sensitivity, regardless of protein aggregation on the surface of the device. It's worth noting that saturation on the surface affects the surface plasmon resonance and not the localized plasmon resonance in the pyramid, as demonstrated in Fig. S4 of the supplementary material.

\section*{Conclusions}

To conclude, this study demonstrates the occurrence of double-anticrossing between two LSPR modes and a single SPP wave in a pyramidal nanoparticles lattice through the variation of the angle of incidence of the input electric field. The geometric composition of these nanoparticles favors this phenomenon due to the induced interaction between the electric field and the pyramidal structure, which increases the density of free charges.

The tunable absorption wavelength by the angle of incidence of the input electric field was demonstrated. Although the previous one is a non-conventional method for observing this effect and the anticrossing above, there are widely demonstrated tuning methods involving varying the elements' size, separation, and shape of the nanoparticles. 

It is important to note that during the anticrossing process, the SPP wave disappears at an incidence angle of \ang{37}. However, it then couples with an LSPR mode that arises when the angle of incidence is \ang{39}. This transition is noticeable in the region where the LSPR mode, located at 850 nm, reaches its maximum intensity.

In our experiments, we demonstrated an increase in the intensity reached by a blue-shifted LSPR mode located in the red wavelength region as a function of the variation of the angle of incidence of the input electric field. The results showed that the electric field intensity achieved by the pyramidal system at an angle of incidence of \ang{0} as a function of the incident electric field intensity was 2.62 times higher. However, by adjusting the angle of incidence at \ang{40}, it is possible to observe that the electric field intensity reached was about 68.10 times higher compared to the pyramidal structure at \ang{0}. This suggests that implementing this structure for SERS applications could lead to better performance at certain angles of incidence of the input electric field.

A nearly constant LSPR mode at 850 nm was demonstrated. This LSPR mode controls and synchronizes the other resonances in the proposed system. The modes that will cause the anticrossing interaction later appear when this mode begins to emerge, and when it reaches its maximum intensity, the transition between the SPP wave and the LSPR mode is observed. We analyze the plasmonic response of this device for different concentrations of the protein Pb27r from \textit{P. brasiliensis}. The studies carried out show that a sensitivity of 1.546 nm/$\mu$g/ml was achieved, which is $10^3$ times greater than that obtained with conventional LSPR systems, demonstrating an excellent performance when implemented in biosensing applications. Moreover, these characteristics also strongly favor the improvement of light capture for photovoltaic systems, and obtaining highly-efficient metamaterials.

\section*{Methods}

\subsection{Numerical modeling}

A regularly spaced array of pyramids has been studied using 3D-Finite Element Method (3D-FEM) simulations in COMSOL Multiphysics. The computational domain was simplified using geometrical symmetry, so only one unit cell of the array needed to be simulated. The unit cell consists of a gold square pyramid in the center with all edges of the same length (L), a gold substrate under the pyramid's base, and air covering the top of the pyramid and the exposed substrate. The structure was excited by a plane wave with varying angles of incidence and polarization. In this manuscript, we investigated the influence of several parameters: pyramid edge length (L), pyramid spacing (D) - the distance between the edges of adjacent pyramids, and angle of incidence of the plane wave excitation. The metric employed to evaluate the designs was the electric field intensity on the nanostructured device surface (both pyramid and substrate), as shown in Fig.~\ref{fig1}b. The boundary conditions were set as follows: Perfect Electric Conductor (PEC) beneath the gold substrate, Periodic Boundary Conditions (PBC) on the four lateral sides of the unit cell, and a port input on the top side of the cell, where the plane wave excitation occurs.    

\subsection{Manufacturing process of pyramidal nanoparticle lattice}
The process of fabricating the pyramidal nanoparticle array comprises several well-defined stages, as illustrated in Fig S1 from the supplementary material. Firstly, a silicon wafer with a (100) crystallographic orientation is carefully chosen for the initial stage, which is motivated by reasons that will be explained later. This chosen wafer is then coated with a uniform 40 nm-thick layer of chromium (Cr) using a thermal evaporation technique. To generate a mask consisting of an array of 150 nm diameter circular windows, the substrate is coated with a layer of PMMA (polymethyl methacrylate) resist using a spin-coating process. This mask is generated with 30 kV Raith e-LiNe Plus electron beam lithography system, which strategically maintains the exposed Cr layer after development. A targeted Cr etching step is undertaken to reveal the silicon substrate beneath the circular lithographic mask, which is then immersed in a potassium hydroxide (KOH) solution. This immersion creates a distinctive pyramidal array structure on the substrate's surface.

Due to the chosen orientation of the wafer and the anisotropic characteristic of the etching process, the resulting structure is an array of pyramidal-shaped cavities under each circular window. The chromium mask can then be removed, and the result corresponds to a silicon mold for the pyramidal nanoparticle array. Next, 175 nm thick layer of gold is thermally evaporated to fill every cavity and cover the silicon substrate. Finally, a glass substrate is glued to the gold. Since gold and silicon do not adhere well, a gold layer conformal to the mold can be extracted. The silicon template remains undamaged in this step, allowing it to be reused in future evaporation steps. The final structure corresponds to a pyramidal nanoparticle array over a gold substrate on a glass substrate, as shown in Fig.~\ref{fig1}c.

\subsection{Optical setup}

Fig. S5 from the supplementary material shows a scheme describing the optical setup used to characterize our devices. The setup comprises a white halogen lamp that serves as a light source and is directed towards a spatial filter consisting of two objective lenses. The first lens has a 0.40 mm aperture and 20x amplification, while the second has a 0.20 mm and 8x amplification. A 100 $\mu$m pinhole is placed between them, which acts as a low-pass filter to reduce noise from the light source. This results in a collimated Gaussian beam focused on the substrate with the pyramidal nanoparticle lattice. The substrate is mounted on an \textit{x-z} translation stage with \textit{y} rotation, enabling the substrate to move and directing the light beam to the desired position. Finally, the reflected beam is focused through a plano-convex lens onto the input of an optical fiber that guides the reflected signal to the CCD Spectrometer.

\subsection{Raman spectroscopy}

carbonyl functional group. It has a high solubility in water, a melting point of 135°C, and is easy to synthesize, making it ideal for experiments. The most cited theoretical study on the Raman spectrum of urea suggests that it has 12 vibrational modes, with the symmetric vibration mode of the C-N chains having the highest intensity of counts in a Raman displacement measurement. An aqueous solution of urea was prepared, and a drop was deposited on a substrate containing nanopyramids and smooth gold regions. The substrate was placed on a hotplate for rapid drying, which prevents material accumulation and ensures homogeneous distribution of the analyte. The measurements were taken using a WiTec Alpha 300 Raman spectrometer with lasers of different wavelengths.

\subsection{Data analysis}

Fig.~\ref{fig1}b was generated using the normalized electrical field intensity obtained from simulations of the pyramidal nanoparticle lattice with a finite element method, while varying the pyramid's side and spacing. Similarly, the electric displacement field distribution obtained from the pyramidal nanoparticle lattice was used to construct Fig.~\ref{fig2}c. Once the devices were manufactured, absorption spectra were measured at various angles of incidence, and the spectral response was analyzed and processed using normalization, detrending, and linear regression techniques to create Fig.~\ref{fig2}a and b. The same procedures were applied to analyze the data obtained from the plasmonic resonance measured as a function of the variation in the concentration of the Pb27r protein, which allowed for the calculation of its corresponding wavelength shift. This information was then used to create Fig.~\ref{fig3}a and b. Finally, the Raman response from the pyramidal lattice and smooth metal surface was measured for different wavelengths and used to build Fig.~\ref{fig2}d.

\subsection{Pb27r protein synthesis and sample preparation}

The recombinant Pb27 protein from the \textit{Paracoccidioides brasiliensis} - Pb18 fungal strain was produced through heterologous expression in Escherichia coli BL21(DE3) and purified using fast protein liquid chromatography, following the method described in~\cite{coutinho2019}. Purification steps were assessed using SDS-PAGE 15\% with Coomassie Blue staining, and protein concentration was determined by UV spectrophotometry. The purified sample (3.36 mg/mL) was stored at -20°C in 20 mM sodium phosphate buffer pH 7.4 with 50 mM NaCl and 10\% glycerol. For plasmonic resonance measurements, Pb27r was diluted to a working concentration of 1 mg/mL in DEPC water. Serial dilutions were then performed to achieve concentrations of 15.62, 10.41, 7.81, 5.20, 3.90, and 1.95 $\mu$g/mL.

\section*{Data availability}
The data that support the findings of this study are available from the corresponding author upon reasonable request.

\section*{Acknowledgments}

The authors acknowledge that this work was supported by CAPES, Finep (SibratecNano, 01.13.0357.00, 0513/22), Fapemig (APQ-01602-21, APQ-02286-23, APQ-05305-23), and CNPq. Partial financial support was also received from the project Omni XGM-AFCCT-2024-3-1-1 supported by xGMobile – EMBRAPII-Inatel Competence Center on 5G and 6G Networks, with financial resources from the PPI IoT/Manufatura 4.0 from MCTI grant number 052/2023, signed with EMBRAPII.

\section*{Author contributions}

\textbf{Theoretical construction and numerical analysis:} YHI, TEMM, PSSG, JRMS, WNR, JCR.
\textbf{Manufacturing of all-metal nanopyramidal metasurface:} FCS, RSG, WNR, JCR.
\textbf{Optical characterization (LSPR and Raman Spectroscopy measurements) and performance analysis as a biosensor:} TEMM, FMFT, FCS, BPD, EMNM, CAF, JCR.
\textbf{Pb27r protein synthesis and sample preparation:} EMNM, BPD, LMA, AR, RAPN, CAF.
\textbf{Data analysis and processing:} TE, YHI, PSSG, JRMS, MGB, JCR

\section*{Competing financial interests}
The authors declare no competing financial interests

\section*{References}
\bibliography{references}


\end{document}


\title{Tunable Surface Plasmon-Polaritons Interaction in All-Metal Pyramidal Metasurfaces: Unveiling Principles and Significance for Biosensing Applications}

\author{Talles E. M. Marques}
\affiliation{\UFMGDELT}

\author{Yuri H. Isayama}
\affiliation{\UFMGLCPN}
\affiliation{\UFMGDF}

\author{Felipe M. F. Teixeira}
\affiliation{\UFMGDELT}
\affiliation{\UFMGPPGEE}

\author{Fabiano C. Santana}
\affiliation{\UFMGPPGF}

\author{Rafael S. Gonçalves}
\affiliation{\UFMGLCPN}
\affiliation{\UIL}

\author{Aline Rocha}
\affiliation{\UFMGICB}

\author{Bruna P. Dias}
\affiliation{\CDTN}

\author{Lidia M. Andrade}
\affiliation{\UFMGDF}

\author{Estefânia M. N. Martins}
\affiliation{\CDTN}

\author{Ronaldo A. P. Nagem}
\affiliation{\UFMGICB}

\author{Clascidia A. Furtado}
\affiliation{\CDTN}

\author{Miguel A. G. Balanta}
\affiliation{\UFU}

\author{Jorge Ricardo Mej\'ia-Salazar}
\affiliation{\Inatel}

\author{Paulo S. S. Guimarães}
\affiliation{\UFMGPPGF}
\affiliation{\UFMGDF}

\author{Wagner N. Rodrigues}
\affiliation{\UFMGPPGF}
\affiliation{\UFMGDF}

\author{Jhonattan C. Ramirez *,}
\affiliation{\UFMGDELT}
\affiliation{\UFMGPPGEE}

\maketitle

\textbf{\large Supplementary Materials}

\textbf{Manufacturing process for the Pyramidal Nanoparticle Lattice}

The process of fabricating the SERS substrate involves several distinct stages, illustrated in Fig. S1. The initial step entails selecting a silicon wafer oriented along the (100) crystallographic plane (see Fig. S1a), a choice driven by reasons elaborated upon later. Subsequently, the wafer undergoes sequential sonication in acetone, isopropyl alcohol (IPA), and deionized (DI) water for five minutes each and then blown with an N$_2$ gun.
This selected wafer is then uniformly coated with a 40 nm-thick layer of 3N purity chromium (Cr) using a thermal evaporation technique (see fig. S1b), sourced from Kurt J. Lesker Company. Following this, a layer of polymethyl methacrylate (PMMA) 950K from Allresist GmbH, type AR-P 672.02, is spin-coated onto the sample at 8000 rpm, forming a 66 nm-thick layer. A subsequent soft bake at 180°C for 60 seconds evaporates the solvent, creating a uniform, thin resist layer.

For e-beam lithography, a 30 kV Raith e-LiNe Plus electron beam lithography system is employed to generate a mask comprising an array of circular windows with a diameter of 150 nm spaced by 100 nm, applying a dose of 125 $\mu$C/cm². To preserve the exposed Cr layer strategically, the sample undergoes cold development with a mixture of IPA and DI water (3:1 ratio) at -10°C for 90 seconds. 

The chromium mask was then removed by immersing the sample in chromium etchant 3Ce(NH$_4$)$_2$(NO$_3$)$_6$ from MicroChemicals GmbH for 180 seconds at room temperature, revealing the silicon substrate beneath the circular lithographic mask (see fig. S1d). Subsequently, the sample is immersed in a potassium hydroxide (KOH) solution, creating a distinctive pyramidal array structure on the substrate's surface.
Due to the wafer's specific orientation and the etching process's nature, each circular window formed an array of pyramidal-shaped cavities on the substrate (Fig. S1e). The chromium mask was then removed by the same chromium etchant mentioned previously, yielding a silicon mold for the SERS substrate (Fig. S1f).

Subsequently, a 175 nm-thick layer of 4N purity gold, also supplied by Kurt J. Lesker Company, was thermally evaporated to fill each cavity and coat the silicon substrate (Fig. S1g). A glass substrate was then adhered to the gold layer using the epoxy Tek bond Araldite Professional, taking advantage of the poor adhesion between gold and silicon. This allowed for the extraction of a gold layer conforming to the mold without damaging the silicon template, enabling its reuse in future evaporation steps. This approach significantly increased device throughput by eliminating the need for complex and time-consuming lithography processes.
The final structure comprised a nano pyramidal lattice on a gold substrate (Fig. S1f) mounted on a glass substrate, as depicted in Figure 2 in the main manuscript. 

\begin{figure}[htp]
  \includegraphics[width=\linewidth]{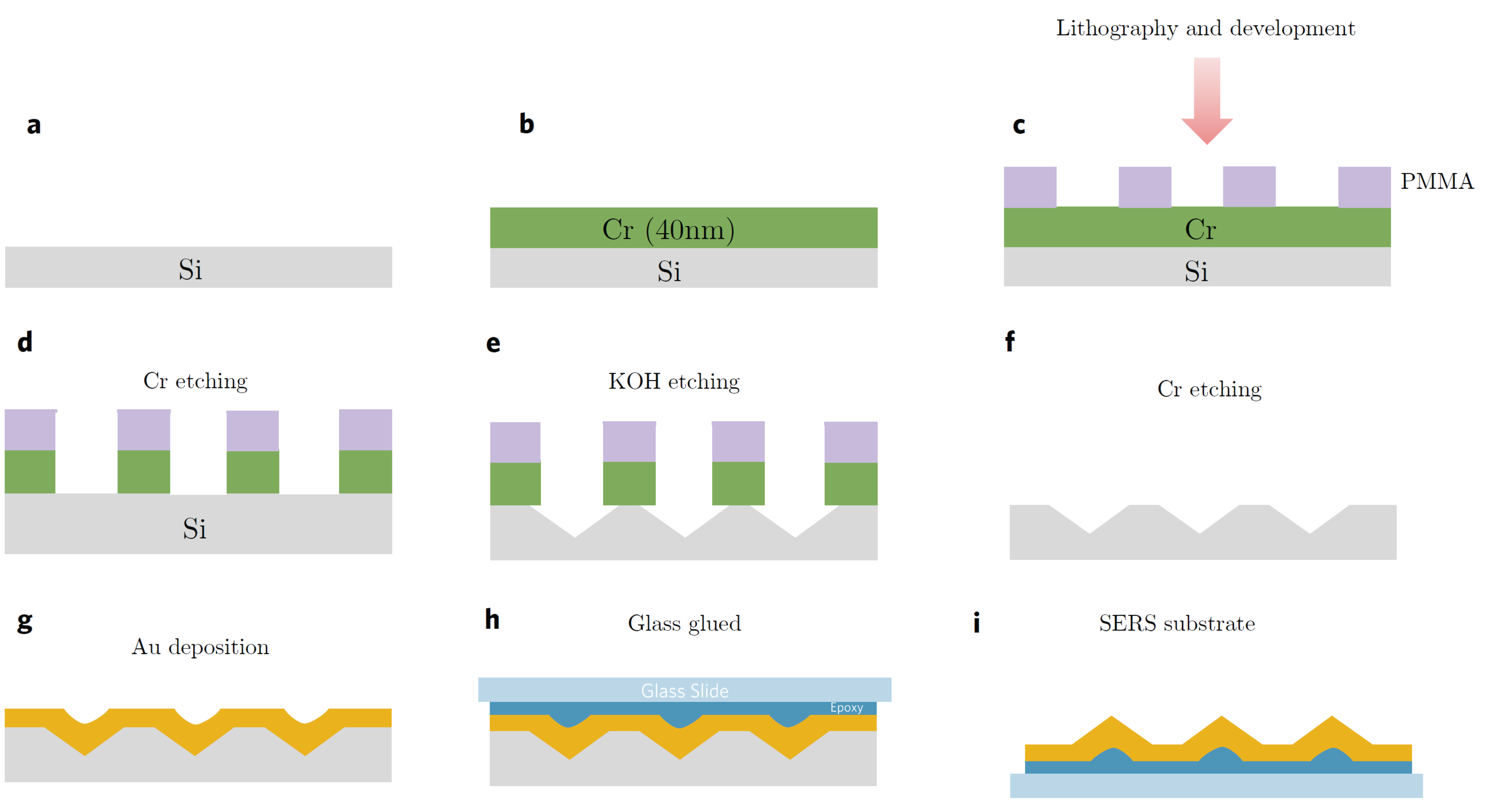}
  \caption{Methodology of the manufacturing process carried out for the pyramidal nanoparticle lattice.}
  \label{figS1}
\end{figure}

\break

\textbf{Electric Field Distribution on All-Metal Pyramidal Metasurface Excited by Plane Waves at Different Angles of Incidence}

\begin{figure}[htp]
  \includegraphics[width=\linewidth]{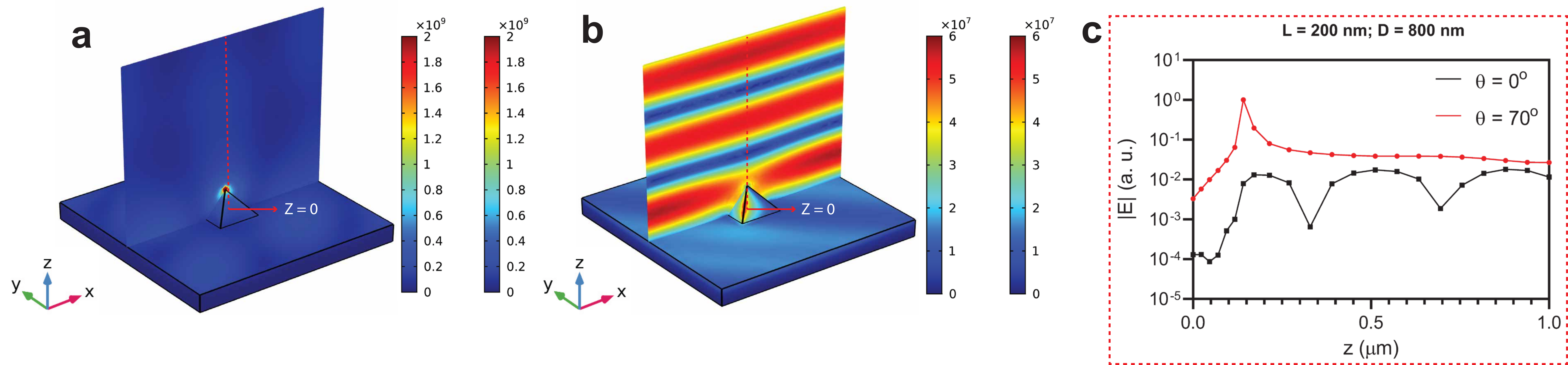}
  \caption{Example of electric field distribution on the pyramids for excitation with plane waves with different angles of incidence. \textbf{(a)} Plane wave excitation with angle of incidence $\theta = 70^o$ and azimuth angle (polarization) $\phi = 45^o$: there is efficient excitation of a plasmonic resonance situated at the pyramid tip, with a very high electric field intensity. \textbf{(b)} Plane wave excitation with angle of incidence $\theta = 0^o$ and azimuth angle (polarization) $\phi = 45^o$: a lot of the incident radiation is reflected, as seen in the standing wave pattern that is formed. Under $\theta = 0^o$, the resonance is formed on the lateral edges of the pyramid, instead of the tip. Furthermore, by employing $\phi = 45^o$ the excited resonance occurs at the edges aligned with the incident field polarization. Electric field intensity on the surface of the pyramid is still high, but two orders of magnitude smaller than the case of oblique incidence. \textbf{(c)} Electric field intensity over the vertical line passing through the tip of the pyramid and orthogonal to the gold substrate plane (represented by the dashed red line in Figs 2a and 2b). The black curve ($\theta = 0^o$) shows the standing wave formed by the almost complete reflection of the incident wave, while the red curve ($\theta = 70^o$) shows the concentration of electric field at the tip of the pyramid and how the resonance is much larger in the case of oblique excitation.}
  \label{figS2}
\end{figure}

\break

\textbf{Optical Characterization of NP-A2 and Structural characterization of NP-A1}

\begin{figure}[htp]
  \includegraphics[width=0.6\linewidth]{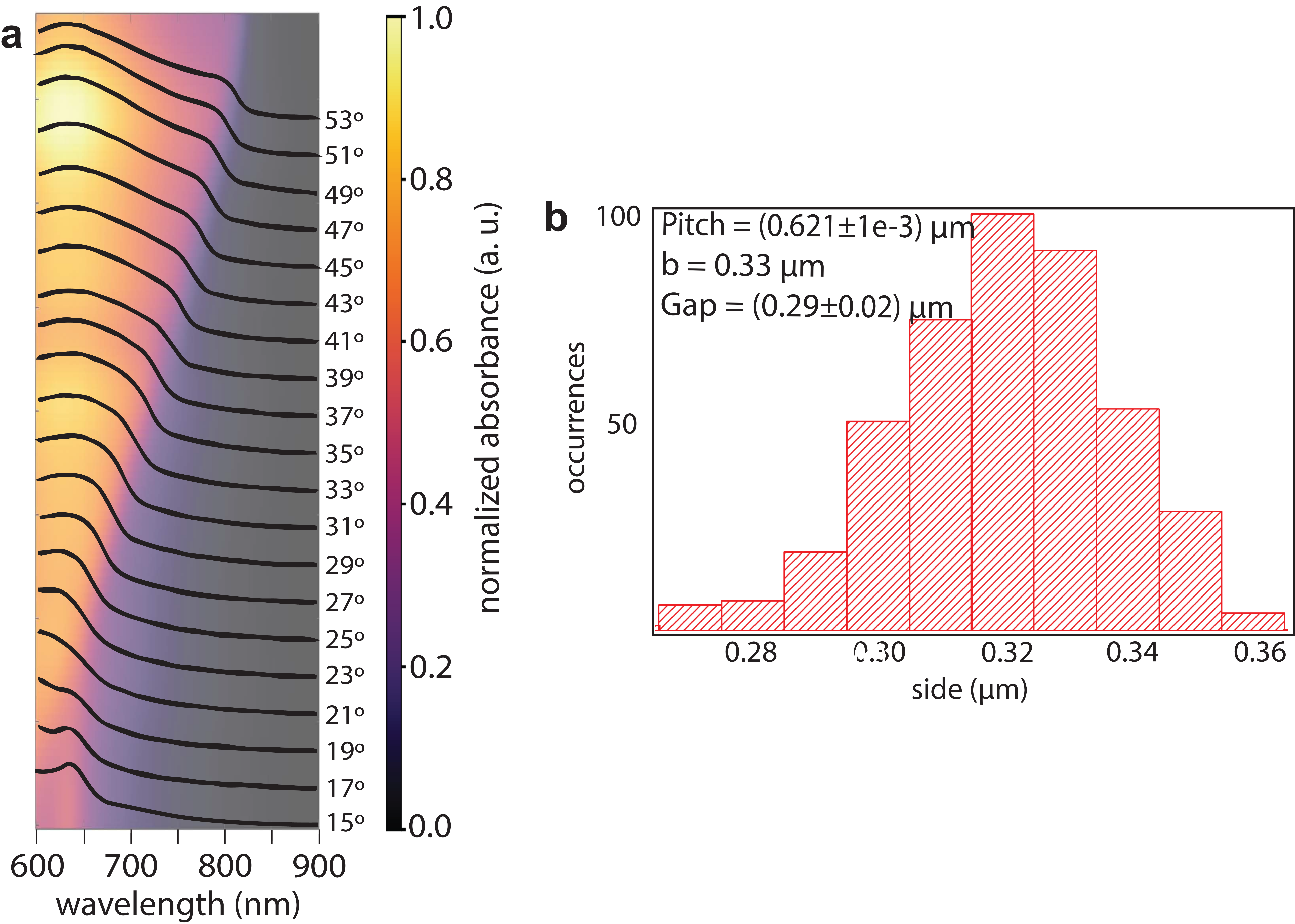}
  \caption{\textbf{(a)} Normalized absorbance for several angles of incidence for an NP-A2 device. From Fig. 1b, we fabricate the NP-A2 devices. This device is located where the electric field intensity reached by the pyramidal nanoparticle lattice is lower than that demonstrated in the isopitch region. For the NP-A2 devices, a slight localized surface plasmon resonance can only be seen at 650 nm when the angle of incidence is \ang{15}; however, this fades as the angle of incidence varies. \textbf{(b)} presents the structural characteristics of the pyramidal nanoparticle lattice corresponding to a NP-A1 device. The NP-A1 devices were designed to have 330 nm of side and 290 nm of gap. The manufactured array has a pitch of (621 ± 1) nm. In addition to the high-quality manufactured pyramidal nanoparticle lattice, as shown in Fig. 1c, it is possible to observe that the designed devices were faithful to the proposed design.}
\label{figS3}
\end{figure}

\break

\textbf{Spectral Response of Pyramidal Nanoparticle Lattice with Albumin Coating and Preparation Details}

\begin{figure}[htp]
  \includegraphics[width=\linewidth]{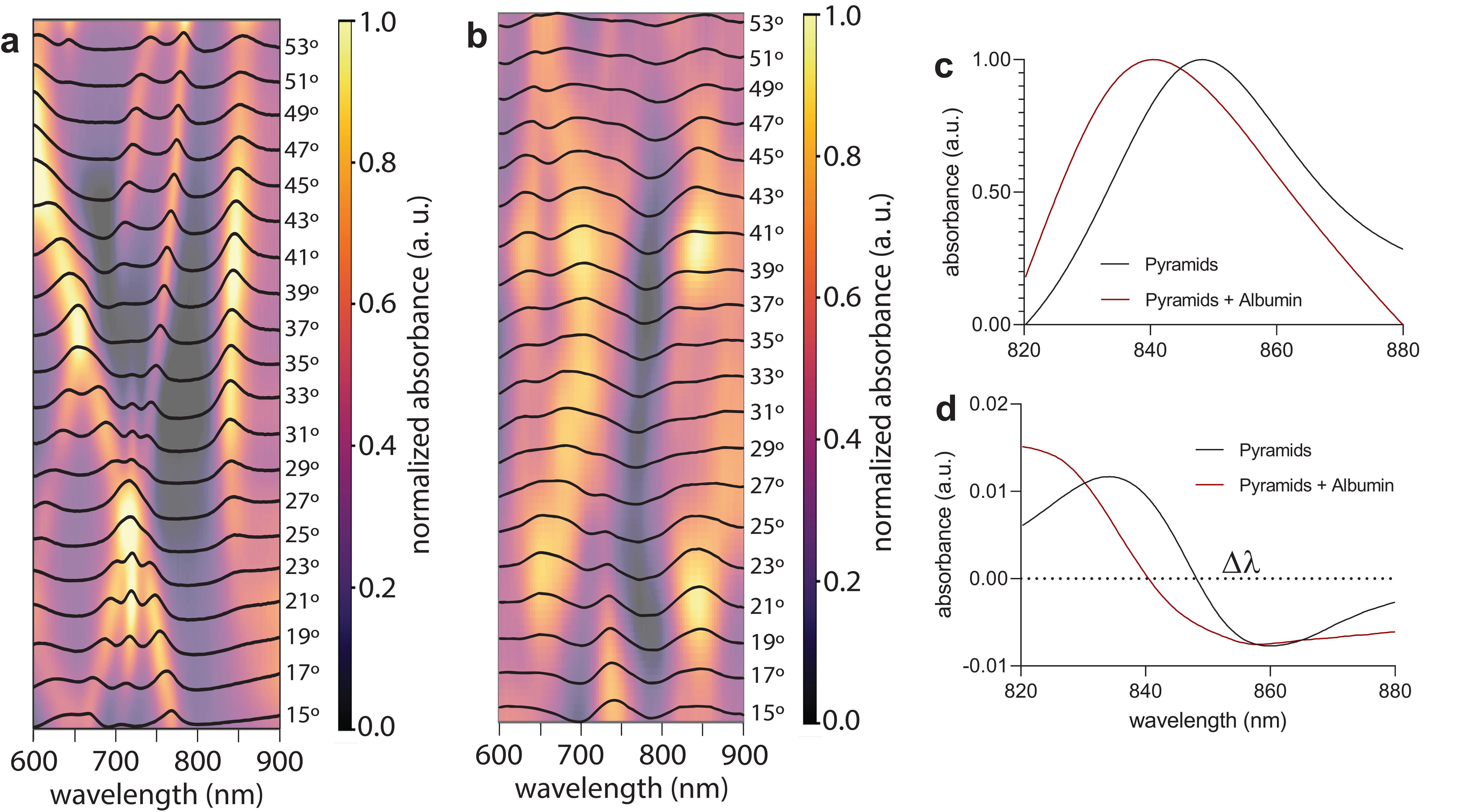}
  \caption{\textbf{(a)} Shows the measured spectral response of the pyramidal nanoparticle lattice. This image was presented originally in the main document; however, it is being included in the supplementary material to make a more direct comparison with the results presented here. \textbf{(b)} Measured spectral response of the pyramidal nanoparticle lattice with albumin on its surface. Unlike a, in b, no SPP is observed; only the LSPRs excited in the said structure. This is because the albumin concentration, i.e., 5 mg/mL, saturates the surface, hindering the excitation of the propagating surface plasmons. \textbf{(c)} Shift in wavelength in the presence of albumin. \textbf{(d)} Presents the displacement detailed in c, but through the derivative of the curves obtained. \textbf{Preparation of the Albumin solution:} 5mg of Albumin fraction V (Merck KGaA, Germany) was dispersed in 1mL of Mili-Q water using an axial shaker. 10µL of this solution was added to the surface of the pyramidal nanoparticles lattice. After drying, the measurements were carried out. }
  \label{figS4}
\end{figure}

\break

\textbf{Schematic Description of the Experimental Setup}

\begin{figure}[htp]
  \includegraphics[width=0.8\linewidth]{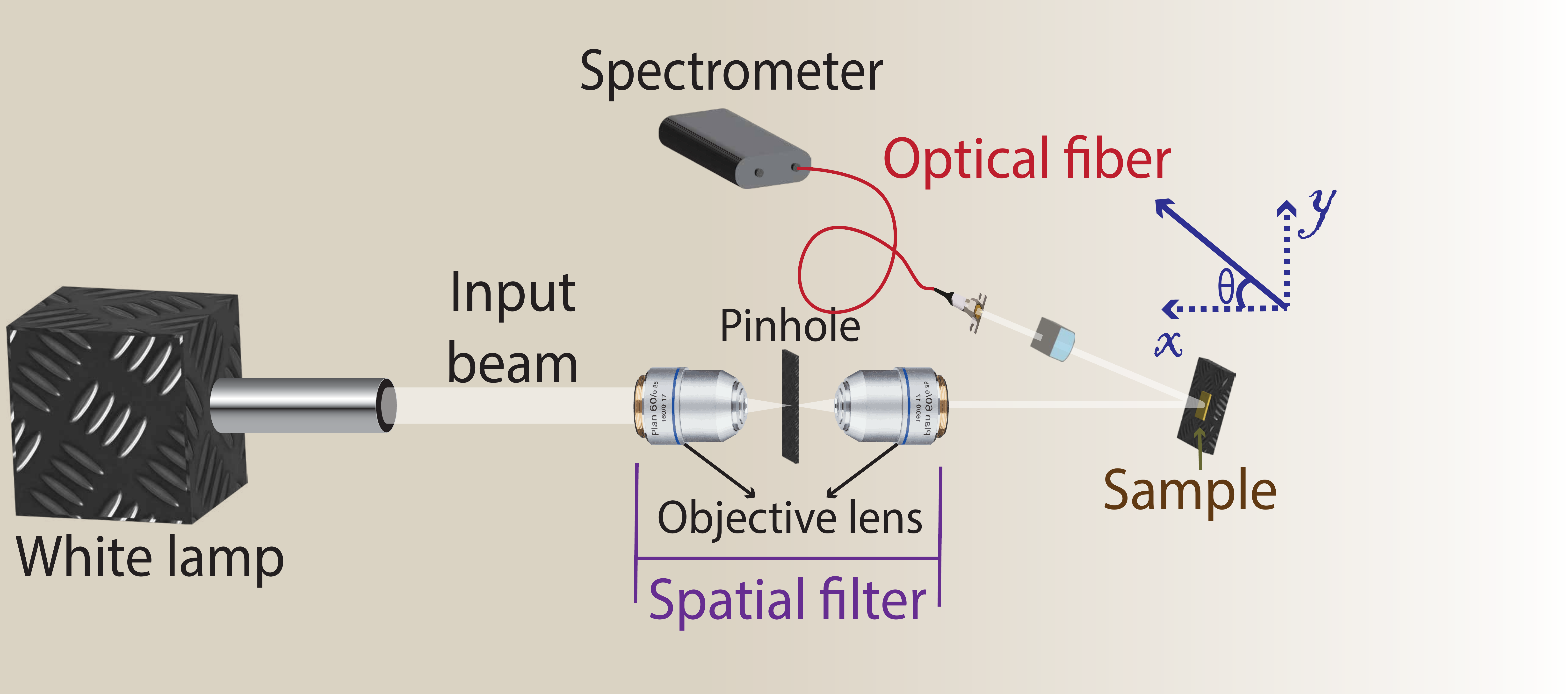}
  \caption{A schematic description of the experimental setup which was implemented to carry out the measurements presented in this manuscript.}
  \label{figS5}
\end{figure}

\break

\textbf{Urea Solution on substrate}

\begin{figure}[htp]
  \includegraphics[width=\linewidth]{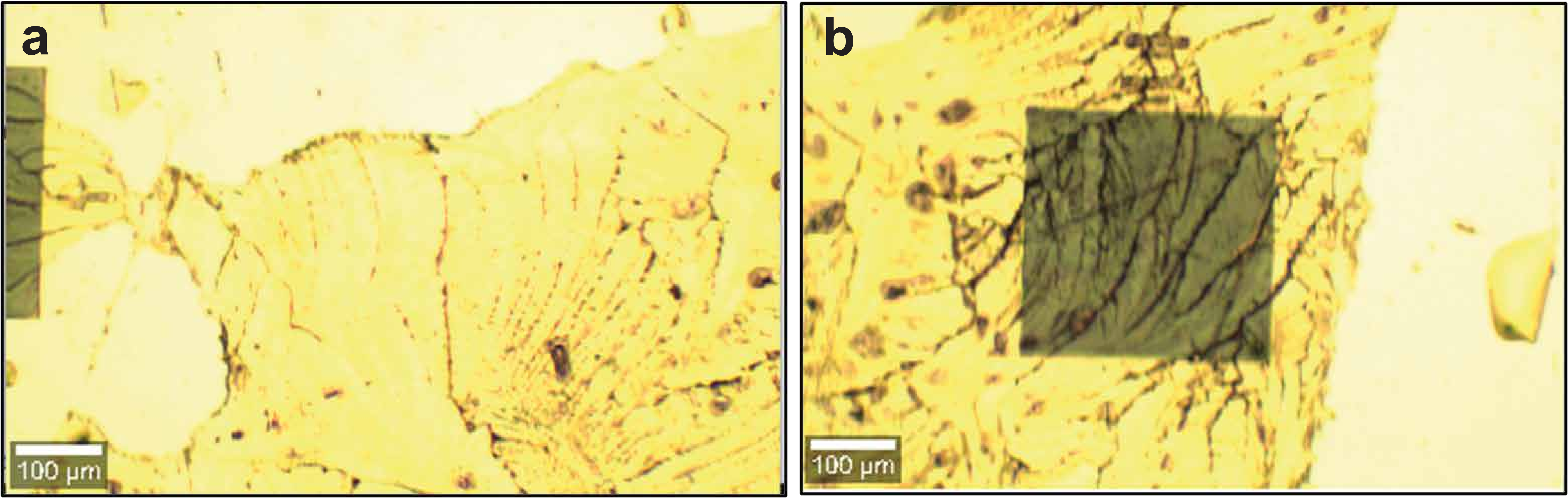}
  \caption{Drops of 40 mMol/L of urea solution deposited on the two regions of the substrate, \textbf{a} Flat gold region, \textbf{b} pyramidal nanoparticle lattice region.}
  \label{figS6}
\end{figure}
---

